\documentclass[aps,prd,reprint,showpacs,floatfix,nofootinbib,a4paper,superscriptaddress]{revtex4-1}

\usepackage[english]{babel}
\usepackage{graphicx}
\usepackage{array}
\usepackage{dcolumn}
\usepackage{bm}
\usepackage{amssymb}
\usepackage{amsmath}
\usepackage{amsfonts}
\usepackage{amsbsy}
\usepackage{color}
\def\scr{$\tilde{\tau}_{\text{CR}}$}

\def\glui{\tilde{g}}
\def\squa{\tilde{q}}

\def\chapm{\chi^{\pm}_1}

\begin{document}

\title{Probing dark matter and CMSSM with same-sign dilepton searches at the LHC}

\author{M.~Cannoni}
\affiliation{Departamento de F\'isica Aplicada, Facultad de Ciencias
Experimentales, Universidad de Huelva, 21071 Huelva, Spain}

\author{O.~Panella}
\affiliation{Istituto Nazionale di Fisica Nucleare, Sezione di Perugia, Via Alessandro Pascoli, 06123, Perugia, Italy}

\author{M.~Pioppi}	
\affiliation{Imperial College, South Kensington Campus, London SW7 2AZ, United Kingdom }

\author{M.~Santoni}
\affiliation{Dipartimento di Fisica, Universit\`a di Perugia, Via Alessandro Pascoli, 06123, Perugia, Italy}



\begin{abstract}

We introduce new observables for the study of the inclusive same sign dileptons production at LHC
which are built out of  ratios of the observed number of same-sign dileptons, 
both with same $N(\ell,\ell)$ and different flavor $N(\ell,\ell')$.
As a case study we apply them to the stau coannihilation region 
of the constrained minimal supersymmetric standard model. 
We show that the new variables depend rather mildly on the center of mass energy and how these can be used to 
constraint the parameter space in the $(m_{1/2},\tan\beta)$ plane. 

\end{abstract}


\maketitle

\paragraph*{\bf Introduction.}

The starting of the LHC era will allow us to finally shed light into the last missing piece of the standard model 
(SM), the Higgs boson, and hopefully to probe the supersymmetric (SUSY) extension of the SM that 
represents the most popular solution to  the hierarchy problem, gauge coupling unification and the nature of dark 
matter. In the minimal supersymmetric standard model (MSSM) with R-parity conservation and in 
the mSUGRA inspired constrained version (CMSSM), the lightest neutralino 
is the lightest SUSY particle (LSP),  neutral and  stable and one of the favored
dark matter (DM) candidates~\cite{bertonerev}.

In the framework of the SM, events in proton-proton ($pp$) collisions with two isolated same-sign  leptons in the 
final state, or same-sign dileptons (SSD), 
are very rare. They may come from double gauge boson production $WZ$, $WW$ and decays, double parton scattering or 
$t\bar{t}W$, the last processes yet to be observed in proton-proton collisions.
This makes this signature very natural to look for new physics. 
Isolated SSD represent also a standard search channel for SUSY models~\cite{SSD}.
The main processes that lead to inclusive final states SSD in proton-proton collisions are 
gluino pair production $pp\to\glui\glui $, gluino-squark associate production $pp \to \glui \squa$ and squark pair 
production $p p \to \squa \squa'$.
Cascade decays from these pairs produce easily $\chi^{\pm}_1\chi^{\pm}_1$, $\chi^{\pm}_1\chi^{0}_2$,
$\chi^{0}_2\chi^{0}_2$ 
that decay and eventually lead to SSD that can be of the same flavor (SFSSD)  
$e^\pm$$e^\pm$, $\mu^\pm\mu^\pm$, 
$\tau^\pm\tau^\pm$ or of different flavor (DFSSD), $e^\pm$$\mu^\pm$, $\mu^\pm\tau^\pm$, $e^\pm\tau^\pm$. 
The signal was searched for by CDF at Tevatron~\cite{CDFlsd} and both
the CMS~\cite{CMSlsd} and ATLAS~\cite{ATLASlsd,ATLASgg+ssd} collaborations have already performed searches for these 
particular class of events in the data sample 
of the final 2011 run with the LHC working at the center of mass energy $\sqrt{s}=7$ TeV 
and a  total integrated luminosity of about 5 fb$^{-1}$. 
No evidence for new physics was found and upper bounds on the number of SSD were used to set constraints on the 
parameter space of the CMSSM.

In this brief report we show that by simply counting the number of SSD pairs 
$N(ee)$, $N(\mu\mu)$, $N(\tau\tau)$, $N(e\mu)$, $N(e,\tau)$, 
$N(\mu\tau)$ it is possible to obtain direct information on fundamental CMSSM parameters like, for example,
$m_{1/2}$ and $\tan\beta$. The proposed variables, that are build from the number of SSD, may give preliminary 
informations on the fundamental parameters of the underlying theory.
Once these are known, one has a guide to which decay chain are likely to show up
in the data yields and thus measure the masses of the SUSY particles. 
As a practical example we study the SSD signature in connection with the so-called stau co-annihilation 
region (\scr) ~of the CMSSM parameter space that is of interest for dark matter searches.

\paragraph*{\bf Observables for the SSD signal.}

Let us consider gluino pair production.
The gluino is a Majorana particle and decays with equal branching fractions ($\cal{B}$) to particles and 
antiparticles; this property allows to have two same sign charginos from the decay chains of the pair of sparticles 
produced in the $pp$ collisions. The two same-sign charginos lead to final states with SSD,
neutrinos and LSP. Under this assumption, we introduce the new observables as follows.
The cross section can be approximated by
$\sigma(pp\to 2\chi_{1}^{+} +X)\simeq \sigma(pp\to \tilde{g}\tilde{g})
\times {\cal B}(\tilde{g}\to q\tilde{q})^2
\times {\cal B}(\tilde{q}\to {q'}\chi_{1}^{+})^2$.
We synthetically call  this cross section $\sigma_{\chi\chi}^{\tilde{g}\tilde{g}}$ and 
the branching ratios corresponding to the various chargino's decay chains leading to a lepton plus 
undetected particles (neutrinos and LSP) plus hadronic jets, $\chi^{\pm}_1 \to \ell +X$ with ${\cal B}_{i,\ell}$,
$\ell =e,\mu,\tau$. For a given integrated luminosity, the number of SFSSD is estimated as
$N(\ell \ell) \propto \sigma_{\chi\chi}^{\tilde{g}\tilde{g}} \times ( \sum_{i} {\cal B}_{i,\ell})^2$,
while the number of DFSSD is instead 
$N(\ell \ell') \propto 2 \sigma_{\chi\chi}^{\tilde{g}\tilde{g}} 
(\sum_{i} {\cal B}_{i,\ell})\times 
(\sum_{i} {\cal B}_{i,\ell'})$. 
The factor 2 takes into account the fact that the leptons come from two identical charginos.
In reason of the expected similar behavior of the first two lepton generations and the peculiar role 
held by the leptons and sleptons of the third family, we consider the ratios:
\begin{eqnarray}
\frac{N(l \tau)}{N(\tau \tau)}=2R, \qquad
\frac{N(l\, l)}{N(\tau \tau)}=R^2
\label{NSF}
\label{eq:1}
\end{eqnarray}
with $l=e,\;\mu$ and 
\begin{equation}
R=\frac{{\sum_{i} {\cal B}_{i,l}}}{{\sum_{i} {\cal B}_{i,\tau}}}.
\label{eq:2}
\end{equation}
We remark that when all contributing mechanisms are considered the total cross section ($\sigma_{\chi\chi}= 
\sigma_{\chi\chi}^{\tilde{g}\tilde{g}}+ \sigma_{\chi\chi}^{\tilde{q}\tilde{g}} + 
\sigma_{\chi\chi}^{\tilde{q}\tilde{q}}\dots $) drops out exactly.
Considering the relations ~\eqref{eq:1}, we finally define the variables:
\begin{eqnarray}
\label{eqN1}
N_1 &=& \frac{1}{2}\,\frac{N(e \tau)+N(\mu \tau)}{N(\tau \tau)}\, \approx 2 R\label{N1},
\label{eq:3}\\
N_2 &=& \frac{1}{4}\,\frac{N(ee)+N(\mu\mu)+N(e\mu)}{N(\tau \tau)}\, \approx R^2\label{N2}.
\label{eq:4}
\end{eqnarray}
We choose the normalization factors $1/2$ and $1/4$ in such a way that $N_1$ and $N_2$ are average
quantities that should take values similar to the basic ratios in Eq.~\eqref{eq:1}. 
By definition, these variables should depend very mildly on the production cross section
and on the center of mass energy of the machine.
If we relax the assumption that the SSD are produced through a pair of charginos $\chi_1^\pm\chi_1^\pm $  the 
above Eq.~\eqref{N1} and Eq.~\eqref{N2} are still valid to a very good degree of approximation although the 
cancellation of the cross section does not apply any longer.

\paragraph*{\bf Dark matter and the CMSSM.}

The CMSSM is specified by assigning the value of the common gauginos mass $m_{1/2}$, the common scalars mass $m_0$,
the common trilinear scalar coupling $A_0$ at a certain unification energy scale. The other fundamental parameter
is $\tan\beta$,  the ratio of the two vacuum expectation values of the Higgs doublet. Further we assume the 
Higgs mixing parameter $\mu$ to be positive.
The \scr~ is one of the  regions~\cite{ellisWMAP} of the parameter space in which the relic density of the neutralino 
as main dark 
matter component is compatible with the WMAP~\cite{wmap} measurement of the cosmic microwave background anisotropies,
$0.096<\Omega h^2 < 0.128$ at 3$\sigma$.  
In the \scr~\cite{stauco}, the neutralino relic density 
is controlled by co-annihilation before the freeze-out between the LSP and the lightest stau, $\tilde{\tau}_1$,
which is the next LSP with a mass splitting $\Delta m=m_{\tilde{\tau}_1} -m_{\chi_0}$ of few GeVs. 

In literature there  exist theoretical studies discussing  signatures of the
\scr ~at LHC. In Refs.~\cite{Arnowitt} the authors show that with the reconstruction of  
the decays $\chi^0_2  \to \tilde{\tau}_1 \tau \to \tau\tau \chi^0_1 $ at the end of the gluinos
and squarks cascade, it is in principle possible to measure the masses of the relevant particles 
and infer the values of the CMSSM parameters. In Ref.~\cite{Godbole} it is argued that the measurement 
of the tau polarization in the $\tilde{\tau}_1 \to \tau \chi^0_1 $ can be useful to measure the mass difference 
$\Delta m$. 
The smallness of $\Delta m$ is an important quantity for the phenomenology at LHC~\cite{Arnowitt}, as it sets the scale 
of the transverse momentum of the leptons at the end of the cascade decays, but also for indirect detection signals
in astrophysical searches with gamma ray, as it is responsible for spectral features at the endpoint 
of the gamma spectrum~\cite{ib} and for the large cross section for the annihilation into a $\tilde{\tau}_1$
pair at the galactic center~\cite{BHsusy}.

\begin{figure}[t!]
\includegraphics*[scale=0.4]{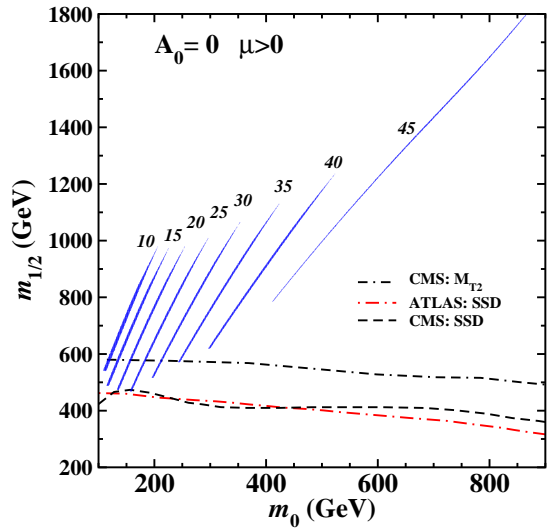}
\caption
{Regions of the CMSSM parameter space in the ($m_0,m_{1/2}$) plane which satisfy the WMAP constraint on the 
relic density  $0.096<\Omega h^2 < 0.128$ for different values of $\tan\beta$ that is given by the numbers above the 
strips.
The dashed,  dot-double-dashed  and dot-dashed lines are respectively the  
the 95\% CL exclusion curves obtained by CMS~\cite{CMSlsd} and 
ATLAS~\cite{ATLASlsd,ATLASgg+ssd} with the SSD search and the CMS SUSY search with hadronic final 
states~\cite{CMSjets}.}
\label{strips}
\end{figure}

We consider the parameter space with fixed trilinear scalar coupling $A_0 =0$,
which is used as benchmark for supersymmetric studies at the LHC. 
The stau-coannihilation strips shown in Figure~\ref{strips} are obtained with 
$\textsf{microOMEGAs}$~\cite{micromegas} using \textsf{SoftSusy}~\cite{Softsusy} as supersymmetric mass spectrum 
calculator, imposing WMAP constraints on the relic density 
accelerator constraints on the lightest Higgs, $m_h >114.4$ GeV,
chargino mass $m_{\chi^{+}_1}>103.5$ GeV and the flavor physics constraint from 
bottom mesons decays $B_s  \to X\gamma$ and $B_s \to \mu^+ \mu^-$.  
In Figure~\ref{strips} we also report the 95\% CL exclusion curves obtained by CMS~\cite{CMSlsd} and 
ATLAS~\cite{ATLASlsd,ATLASgg+ssd} with the SSD search and the CMS SUSY search with hadronic final 
states~\cite{CMSjets}.
As can be seen the first two years of operation of the LHC could only marginally
exclude the parameter space of interest in this work. 
\begin{figure*}[t!]\
\includegraphics*[scale=0.6]{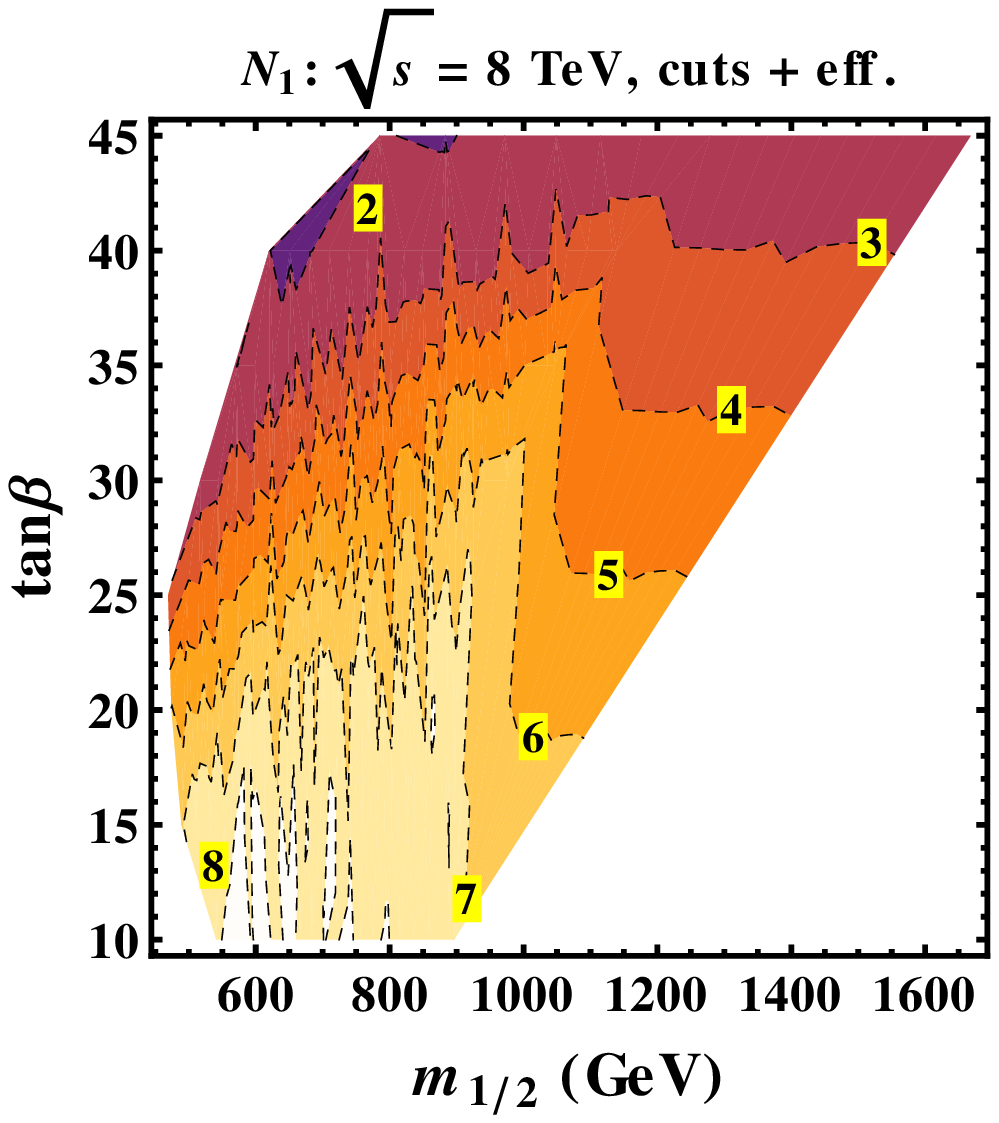}\hspace{1cm}
\includegraphics*[scale=0.6]{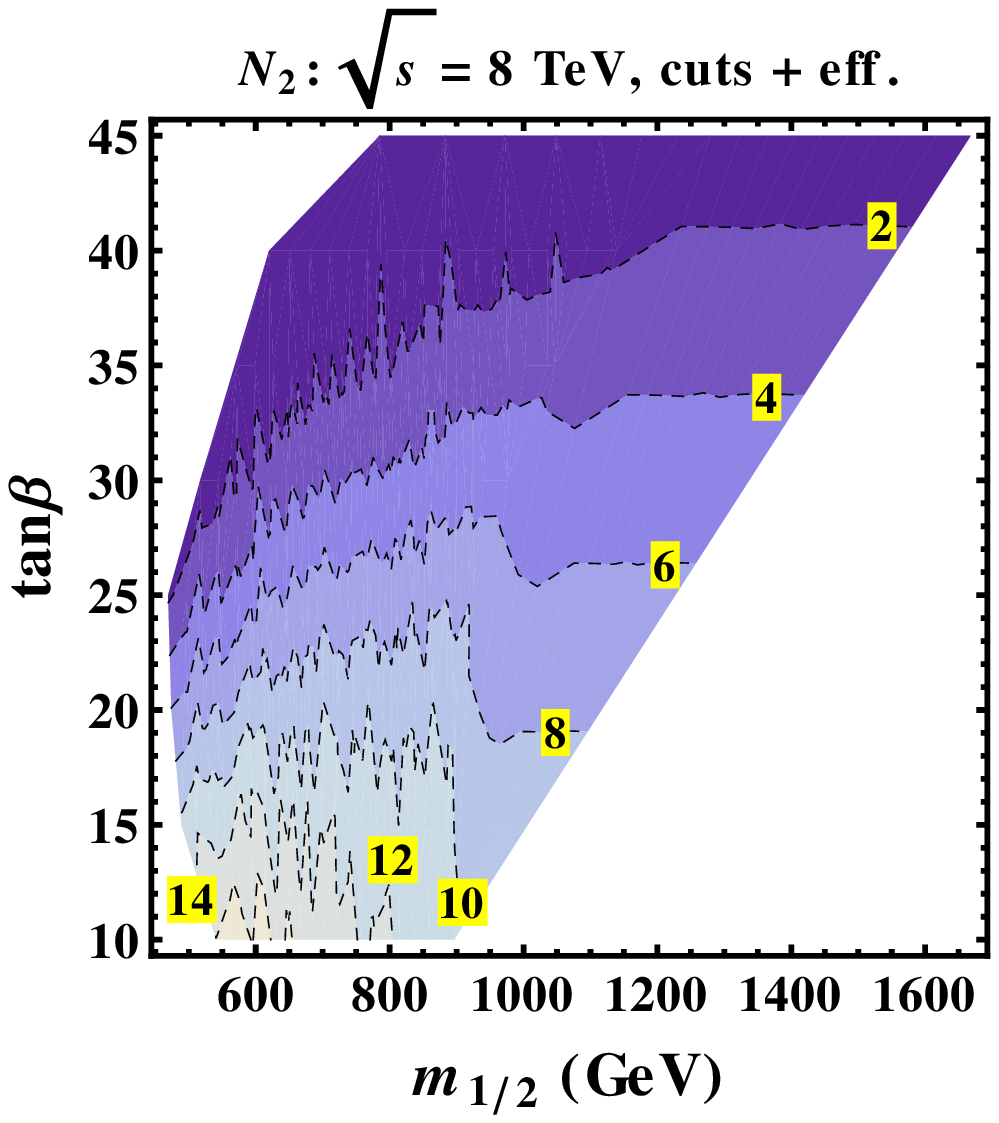}
\includegraphics*[scale=0.6]{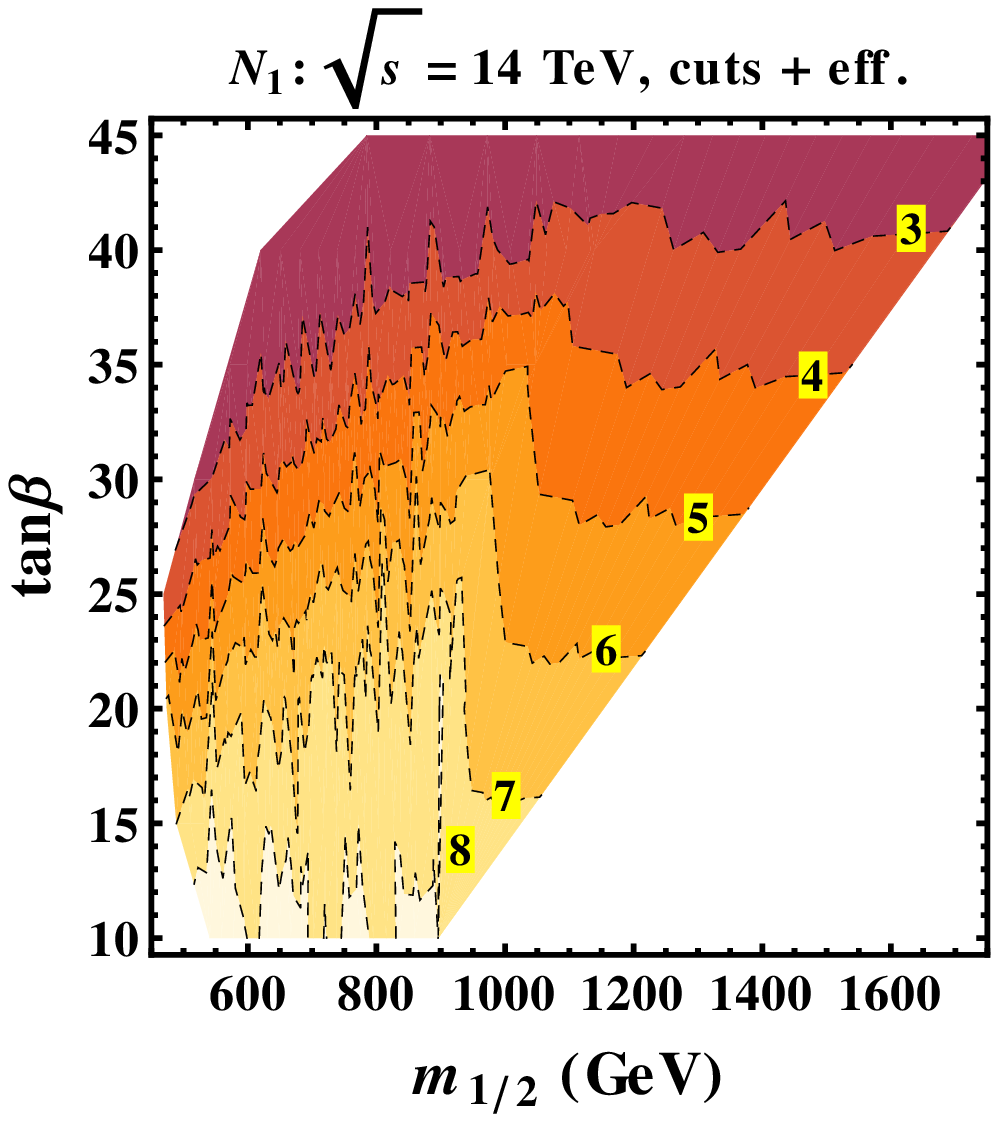}\hspace{1cm}
\includegraphics*[scale=0.6]{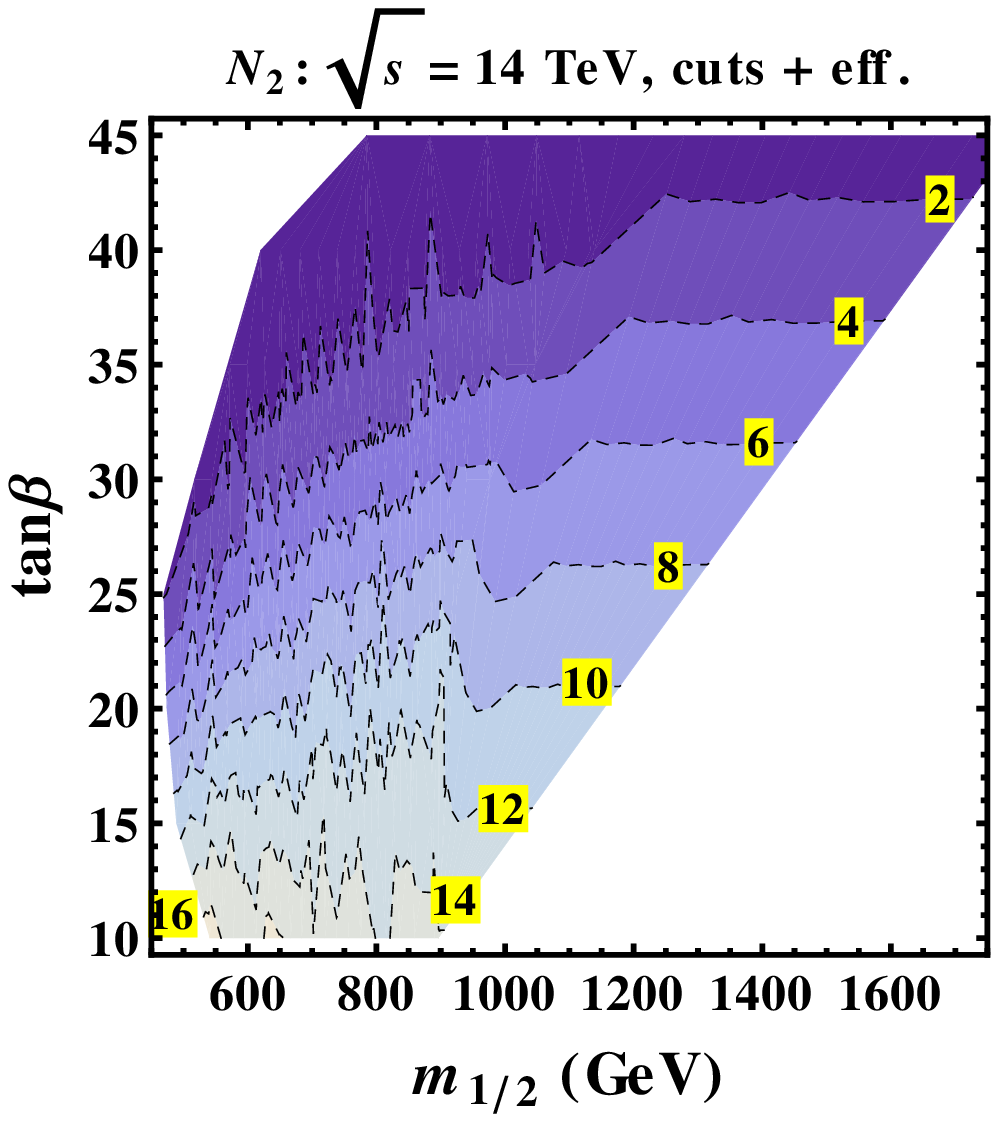}
\caption{Contour maps in the plane ($m_{1/2}, \tan\beta$) of the observables $N_1$ 
(left column) and $N_2$ (right column) as defined in respectively in Eq.~\eqref{eq:3} and Eq.~\eqref{eq:4}. 
Top plots are obtained with a simulation with $\sqrt{s}=8$ TeV, the bottom plots with $\sqrt{s}=14$ TeV.
The cuts and efficiencies are discussed in the text.}
\label{fig:2}
\end{figure*}

\paragraph*{\bf Analysis.} 
To check the validity of the new observables  we make a preliminary simulation selecting 
points along the strips with $\tan\beta =10$ and $\tan\beta =40$.
The theoretical ratio $R$ defined in Eq.~\eqref{eq:2} is calculated in each point using  
\textsf{SusyHIT}~\cite{susyhit} to compute the branching fractions.
All informations of the selected models are then 
passed to \textsf{Pythia 8.1}~\cite{pythia} to generate events in p-\!p collisions at $\sqrt{s}= 14$ TeV.
We have generated $4\times 10^6$ events for each CMSSM point requiring that in the
final state there are same-sign leptons, jets and missing energy.
We find that Eqs.~\eqref{eq:3} and \eqref{eq:4} are well satisfied,
the number of SSD coming from two same-sign $\chapm$ 
is correctly predicted in terms of the theoretical ratio of the chargino's branching ratios $R$.
When all production mechanisms are allowed, and especially at large $\tan\beta$,
the number of taus SSD is contaminated by the decay chain involving the second neutralino,
$\chi^{\pm}_1 \chi^0_2$ and $\chi^0_2 \chi^0_2$ thus, depending on the point of the parameter space, 
deviations up to $50\%$ are observed.

We now want to relate $N_{1,2}$ to $m_{1/2}$ and $\tan\beta$, the two CMSSM parameters that  are the more 
interesting from the dark matter phenomenology point of view. In fact in Ref.~\cite{MCdirect} it was found that the 
neutralino mass along the strips of Fig.~\ref{strips} is roughly given by $m_{\chi}=0.44\times m_{1/2} 
-16\;\text{GeV}$ for all the values
of $\tan\beta$. Furthermore the spin-independent neutralino-nucleon cross section, and hence direct detection rates,
strongly depend on $\tan\beta$. 
\begin{figure*}[t!]
\includegraphics*[scale=0.6]{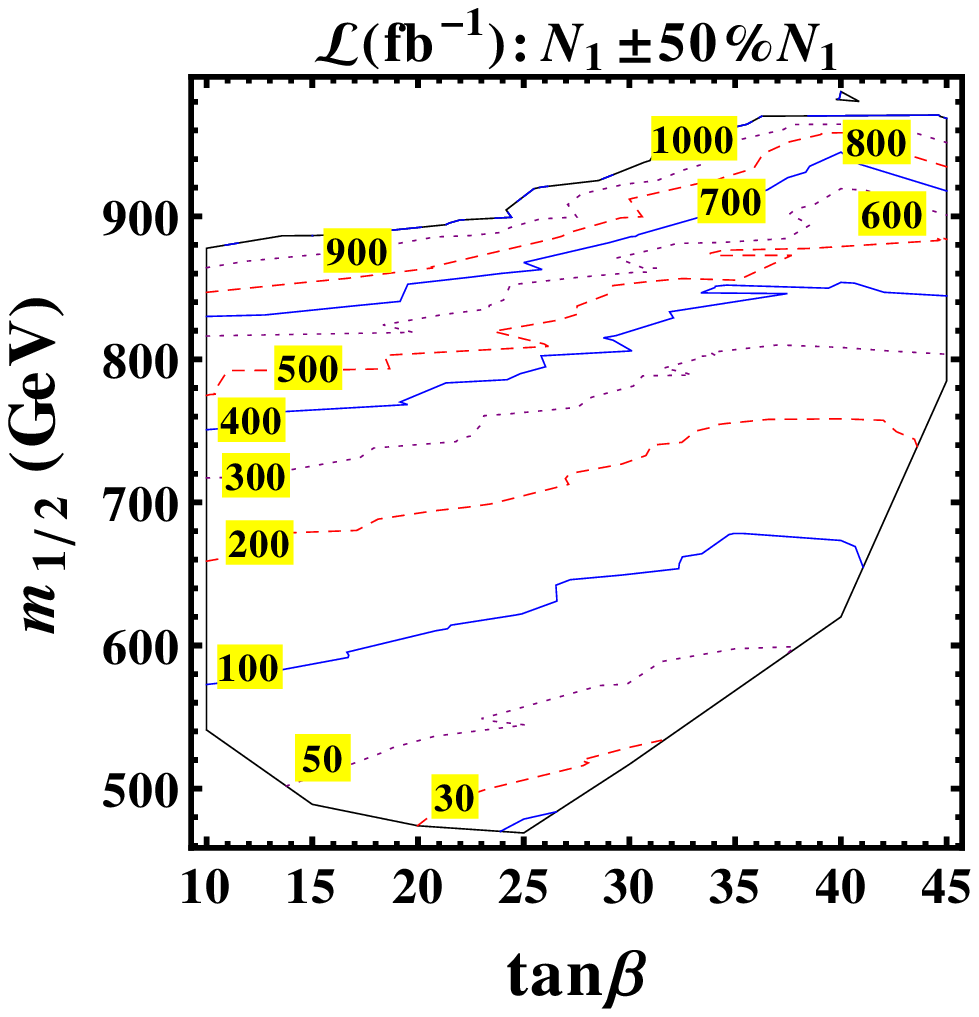}\hspace{1cm}
\includegraphics*[scale=0.6]{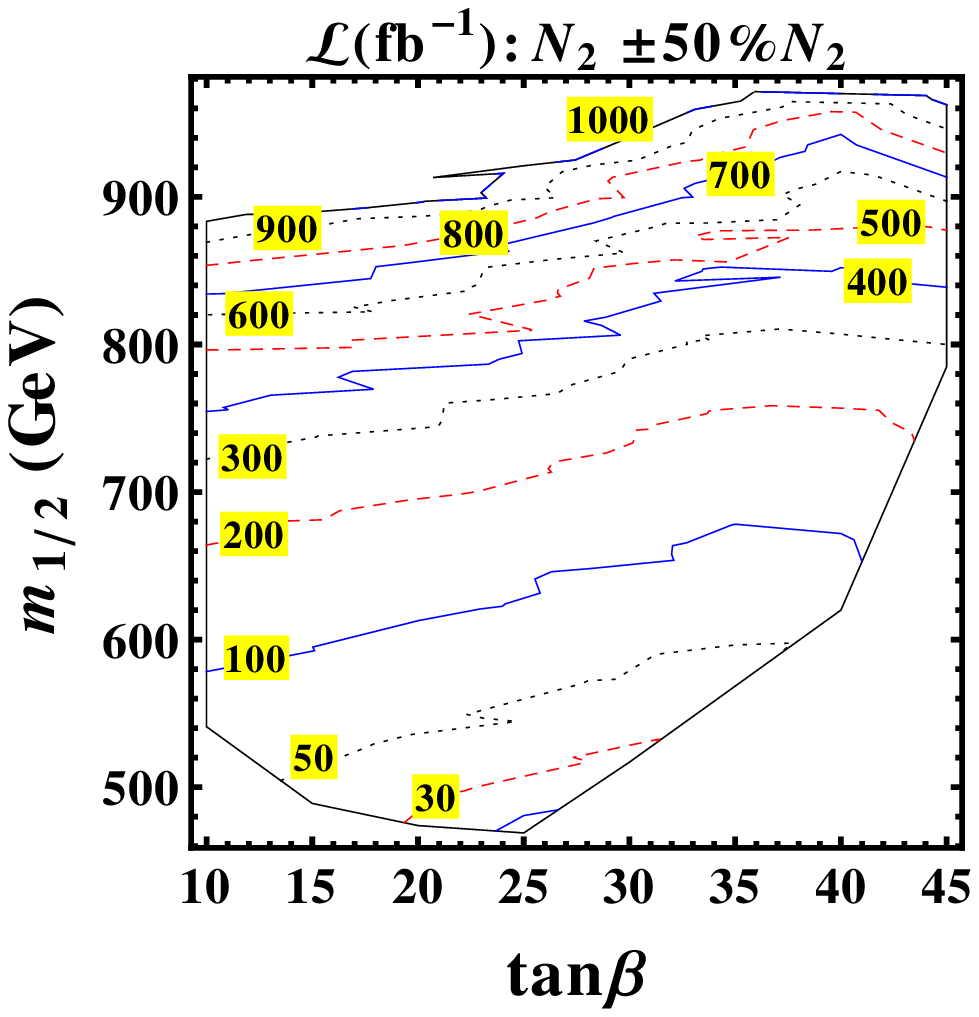}
\caption{Curves in the ($m_{1/2},\tan\beta$) plane showing  the  luminosity in fb$^{-1}$ (boxed values) necessary to 
measure the observables $N_1$ (left panel) and $N_2$ (right panel) with $\pm$ 50\% at the $\sqrt{s}=14$ TeV run, given 
the CMS acceptance cuts and efficiency model discussed in the text. }
\label{Fig:lum14}
\end{figure*}%

In order to cover the full parameter space we took 20 equally spaced points (in $m_{1/2}$) along each strip of 
Fig.~\ref{strips} and  the corresponding value of $m_0$ is chosen to be the center value of the strip width 
along the $m_0$ direction. We  carry out the simulation with 4$\times 10^6$ events for each model point. 
For realistic results we employ the efficiency model for the SSD signal developed by the CMS collaboration
that allows to obtain realistic results at the generator level bypassing the full detector 
simulation~\cite{CMSlsd2,CMSlsd}.
Tau leptons are identified by their hadronic decays, $\tau_h$. 
We imposed all the observed leptons $e$, $\mu$, $\tau_h$ to have a transverse momentum larger than 15 GeV, 
and to be within the acceptance of the ATLAS and CMS detectors ($|\eta|<2.4$). 
The search region to detect experimentally SSD events is defined by two additional variables: 
the transverse hadronic energy $H_T=\sum_{jet} 
{p_T}_{jet}$, determined by all the quarks and gluons with $p_T>40$ GeV within the detector acceptance
and $E^{\text{miss}}_T$, the missing transverse energy determined by all the undetectable particles (neutrinos and 
LSP) in the event and the visible particles outside the detector acceptance.
On top of the lepton selection we require $ H_T >450$ GeV and $E^{\text{miss}}_T$$>120$ GeV. 
The SSD detection efficiency is simulated using the formulas given in~\cite{CMSlsd2,CMSlsd}, 
where the probability for an event to pass the selection cuts is given as a function of 
the $H_T$, $E^{\text{miss}}_T$ and the $p_T$ of the two same sign leptons.
Events with three or more selected leptons are discarded.
Some of the SSD events may even originate from a multi-lepton final state, typically with 3 or 4 leptons, 
where only two of them satisfy the selection criteria. 
The introduction of the kinematic cuts (especially those on $p_T$) lessen the accuracy of the 
approximations in Eqs.~\eqref{N1}~\eqref{N2}, which  have however  been verified to hold when no cuts are applied.
We emphasize that the results of the simulation presented here are obtained considering all the production mechanisms 
and without imposing any selection on the decay chain that lead to SSD.
%

We thus build the contour maps of $N_1$ and $N_2$ in the  
($m_{1/2}, \tan\beta$) plane that  are shown in left panels of Fig.~\ref{fig:2}, $\sqrt{s}=8$ TeV, 
and right panels, $\sqrt{s}=14$ TeV. 
Fig.~\ref{Fig:lum14} shows the luminosity  necessary to achieve an accuracy of 50\% in the measurement of the 
observables $N_1$ (left panel) of $N_2$(right panel) at $\sqrt{s}=14$ TeV. We consider here only the statistical 
error. Systematic errors can be neglected because $N_{1,2}$ are ratios of
observed yields measured using the same selection criteria and obtained in the same experimental conditions.
With the $\sqrt{s}=14 $ TeV run that will follow  after the LHC upgrade, with the possibility of 
accumulating up to 100 - 1000 fb$^{-1}$ of luminosity, will offer the possibility to probe large portions of the 
parameter space in the plane $(m_{1/2},\tan\beta)$ up to values of $m_{1/2}\approx 900 $ GeV.
We do not show the corresponding of Fig.~\ref{Fig:lum14} for $\sqrt{s}=8$ TeV because the production cross sections
are smaller and the planned 20 fb$^{-1}$ are not enough to get $N_{1,2}$ with the same accuracy. 
The rather small differences that can be seen in Fig.~\ref{fig:2} in the values that  $N_{1,2}$ take at the two center 
of mass energies and in Fig.~\ref{Fig:lum14} in luminosity curves, 
confirm our expectation that to a first approximation the observables $N_{1,2}$ are 
independent of the center of mass energy of the collider.
%

\paragraph*{\bf Comments and summary.}
Both the CMS and ATLAS collaborations  have released
results that hint for a possible evidence of a Higgs with mass around 124-126 GeV
in the first $\sim 5$ fb$^{-1}$ of data obtained with $\sqrt{s}=7$ TeV~\cite{CMShiggs,ATLAShiggs}. 
The light Higgs mass in the MSSM receives a large contribution from radiative corrections thus represents 
a crucial quantity to test any SUSY model. 
Updated analysis of the CMSSM parameter space including the new Higgs data~\cite{ellisolive} 
show that $A_0 \neq 0$, large $\tan\beta$ and heavy SUSY spectrum are now generally preferred. 
In the case of the confirmation of the discovery it would be interesting to make detailed study 
of the SSD yield expected from the smaller parameter space, compared to the one explored here, 
compatible with that Higgs mass.  

In our analysis we have excluded those model points found in the right edge 
of the strips at high $m_{1/2}$ in 
which the mass difference between the $\tilde{\tau}_1$ and the LSP ($\tilde{\chi}^0_1$) is less then the mass 
of the tau lepton ($m_\tau \approx 1.7$ GeV).
In this case the two-body decay $\tilde{\tau}_1 \to \tilde{\chi}^0_1 \tau$ is forbidden. 
The $\tilde{\tau}_1$ decays only into suppressed three body final states.
and is a long lived charged particle that have been proposed as a solution to the lithium problem~\cite{LLS2}.
Once produced they can decay outside the detector~\cite{LLS1}.
The sudden reduction of the number of tau SSD $N(\tau\tau)$ is reflected in much higher numerical 
values of $N_1$ and $N_2$  (in particular those of $N_2$). 
For a recent study connecting long-lived staus and SSD in models with the gravitino as the
dark matter candidate see Ref.~\cite{masip}. 

In summary, we have shown that if an excess of SSD relative to the SM yield is observed then  
a measurement of the proposed observables $N_{1,2}$ within a given accuracy  will allow to pin down a given portion of 
the stau co-annihilation region (\scr) of the  CMSSM parameter space in the $(m_{1/2},\tan\beta)$ plane.
This  will in turn also give access to other informations. 
For example if we restrict the possible values of $m_{1/2}$ to a certain range then this will of course also restrict 
the possible values of $m_0$ since these two parameters are related according to the WMAP strips of 
Fig.~\ref{strips}.  
The proposed method can evidently be extended to all SUSY models  
predicting events with final states containing SSD or to different extensions of the standard model that contain 
Majorana particles such as models with weak scale heavy Majorana neutrinos~\cite{HMN}.

\paragraph*{\bf Acknowledgements.}
The work of M.~C. is supported by a MultiDark  under Grant No. CSD2009-00064 of the
Spanish MICINN  Consolider-Ingenio 2010 Program. Further support is provided by the
MICINN project FPA2011-23781.
M.~C., O.~P. and M.~S.~ acknowledge partial support from the Grant MICINN-INFN(PG21)AIC-D-2011-0724. 
Part of this work was done as partial fulfillment for the requirements of the master thesis 
(Laurea Magistrale) of M. ~S. (University of Perugia, December 2011). 

\paragraph*{\bf Note added.}
After submitting this paper for publication both the ATLAS and CMS experiments have reported new evidence at 5$\sigma$ level 
for a scalar particle of mass around $125$ GeV compatible with the Higgs boson~\cite{5s}. 
As already discussed above, and  in view of the newly reported experimental evidence on the Higgs mass, the choice of the trilinear scalar parameter $A_0=0$ 
is not allowed any longer since the radiative corrections needed to achieve $m_h=125$ GeV are driven by the couplings of the third generation sfermions and 
especially so the Higgs-stop-stop trilinear coupling $A_t$ must be large, implying large and negative values for $A_0$. 
The allowed parameter space of the CMSSM in light of the Higgs discovery
is being still more severely constrained and according to some authors this 
model is already disfavored by the data, while others authors do not agree with this conclusion. 
We refer the reader to  Refs.~\cite{ellisolive,CMSSMscan} for a list of recent studies
showing these different points of view.



\begin{thebibliography}{100}

\bibitem{bertonerev}
G.~Bertone, D.~Hooper and J.~Silk,
Phys.\ Rept.\  {\bf 405}, 279 (2005).

\bibitem{SSD}
H.~Baer, X.~Tata and J.~Woodside,
Phys.\ Rev.\ D {\bf 45}, 142 (1992);
R.~M.~Barnett, J.~F.~Gunion and H.~E.~Haber,
Phys.\ Lett.\ B {\bf 315}, 349 (1993).
%
M.~Guchait and D.~P.~Roy,
Phys.\ Rev.\ D {\bf 52}, 133 (1995).
%
H.~Baer, C.~-h.~Chen, F.~Paige and X.~Tata,
Phys.\ Rev.\ D {\bf 53}, 6241 (1996).
%
J.~Alwall, D.~Rainwater and T.~Plehn,
Phys.\ Rev.\ D {\bf 76}, 055006 (2007).
%
H.~Baer, A.~Lessa and H.~Summy,
Phys.\ Lett.\ B {\bf 674}, 49 (2009).
%
K.~T.~Matchev, F.~Moortgat, L.~Pape and M.~Park,
Phys.\ Rev.\ D {\bf 82}, 077701 (2010).


\bibitem{CDFlsd}
A.~Abulencia {\it et al.}  [CDF Collaboration],
Phys.\ Rev.\ Lett.\  {\bf 98}, 221803 (2007).


\bibitem{CMSlsd}
S.~Chatrchyan {\it et al.}  [CMS Collaboration],
arXiv:1205.6615.


\bibitem{ATLASlsd} 
G.~Aad {\it et al.}  [ATLAS Collaboration],
Phys.\ Lett.\ B {\bf 709}, 137 (2012).


\bibitem{ATLASgg+ssd} 
G.~Aad {\it et al.}  [ATLAS Collaboration],
arXiv:1203.5763.



\bibitem{ellisWMAP}
J.~R.~Ellis, K.~A.~Olive, Y.~Santoso and V.~C.~Spanos,
Phys.\ Lett.\ B {\bf 565}, 176 (2003).



\bibitem{wmap}
D.~Larson {\it et al.},
Astrophys.\ J.\ Suppl.\  {\bf 192}, 16 (2011).



\bibitem{stauco} 
J.~R.~Ellis, T.~Falk, K.~A.~Olive and M.~Srednicki,
Astropart.\ Phys.\  {\bf 13}, 181 (2000)
[Erratum-ibid.\  {\bf 15}, 413 (2001)];
M.~E.~Gomez, G.~Lazarides and C.~Pallis,
Phys.\ Rev.\ D {\bf 61}, 123512 (2000);
T.~Nihei, L.~Roszkowski and R.~Ruiz de Austri,
J.\ High \ Energy \ Phys.\  07 (2002) 024;
J.~Edsjo, M.~Schelke, P.~Ullio and P.~Gondolo,
J.\ Cosmol. \ Astropart. \ Phys.\ 04 (2003) 001.


\bibitem{Arnowitt} 
R.~L.~Arnowitt, B.~Dutta, A.~Gurrola, T.~Kamon, A.~Krislock and D.~Toback,
Phys.\ Rev.\ Lett.\  {\bf 100}, 231802 (2008);
%
R.~L.~Arnowitt, A.~Aurisano, B.~Dutta, T.~Kamon, N.~Kolev, P.~Simeon, D.~A.~Toback and P.~Wagner,
Phys.\ Lett.\ B {\bf 649}, 73 (2007);
%
R.~L.~Arnowitt, B.~Dutta, T.~Kamon, N.~Kolev and D.~A.~Toback,
Phys.\ Lett.\ B {\bf 639}, 46 (2006).

\bibitem{Godbole}
R.~M.~Godbole, M.~Guchait and D.~P.~Roy,
Phys.\ Rev.\ D {\bf 79}, 095015 (2009).


\bibitem{ib} 
T.~Bringmann, L.~Bergstrom and J.~Edsjo,
J.\ High \ Energy \ Phys.\ 01 (2008) 049;
M.~Cannoni, M.~E.~Gomez, M.~A.~Sanchez-Conde, F.~Prada and O.~Panella,
Phys.\ Rev.\ D {\bf 81}, 107303 (2010).


\bibitem{BHsusy}
M.~Cannoni, M.~E.~Gomez, M.~A.~Perez-Garcia and J.~D.~Vergados,
Phys. \ Rev.\ D {\bf 85}, 115015 (2012).



\bibitem{micromegas}
G.~Belanger, F.~Boudjema, P.~Brun, A.~Pukhov, S.~Rosier-Lees, P.~Salati and A.~Semenov,
Comput.\ Phys.\ Commun.\  {\bf 182}, 842 (2011).

\bibitem{Softsusy} 
B.~C.~Allanach,
Comput.\ Phys.\ Commun.\  {\bf 143}, 305 (2002).


\bibitem{CMSjets}
CMS-PAS-SUS-12-002,
https://twiki.cern.ch/twiki/bin/\-view/\-CMSPublic\-/PhysicsResultsSUS12002


\bibitem{susyhit}
A.~Djouadi, M.~M.~Muhlleitner and M.~Spira,
Acta Phys.\ Polon.\ B {\bf 38}, 635 (2007)
%
%
M.~Muhlleitner, A.~Djouadi and Y.~Mambrini,
Comput.\ Phys.\ Commun.\  {\bf 168}, 46 (2005)
%
%
A. Djouadi, J. Kalinowski, M. Spira, 
Comput. \ Phys. \ Commun.\ {\bf 108}, 56 (2005). 


\bibitem{pythia}
http://home.thep.lu.se/~torbjorn/Pythia.html

\bibitem{CMSlsd2}
S.~Chatrchyan {\it et al.}  [CMS Collaboration],
J.\ High \ Energy \ Phys.\  06 (2011) 077;
CMS Collaboration,
CMS-PAS-SUS-11-010.


\bibitem{MCdirect}
M.~Cannoni,
Phys.\ Rev.\ D {\bf 84}, 095017 (2011).


\bibitem{CMShiggs} 
S.~Chatrchyan {\it et al.}  [CMS Collaboration],
Phys.\ Lett.\ B {\bf 710}, 26 (2012).

\bibitem{ATLAShiggs}
G.~Aad {\it et al.}  [ATLAS Collaboration],
Phys.\ Lett.\ B {\bf 710}, 49 (2012).





\bibitem{ellisolive} 
J.~Ellis and K.~A.~Olive,
Eur.\ Phys.\ J.\ C {\bf 72}, 2005 (2012).

\bibitem{LLS2} 
T.~Jittoh, K.~Kohri, M.~Koike, J.~Sato, T.~Shimomura and M.~Yamanaka,
Phys.\ Rev.\ D {\bf 76}, 125023 (2007).


\bibitem{LLS1} 
T.~Ito, K.~Nakaji and S.~Shirai,
Phys.\ Lett.\ B {\bf 706}, 314 (2012).

\bibitem{masip} 
R.~Barcelo, J.~I.~Illana, M.~Masip, A.~Prado and P.~Sanchez-Puertas,
arXiv:1206.5108.
  
\bibitem{HMN} 
O.~Panella, M.~Cannoni, C.~Carimalo and Y.~N.~Srivastava,
Phys.\ Rev.\ D {\bf 65}, 035005 (2002).

\bibitem{5s}
CMS-PAS-HIG-12-020, \\
http://cdsweb.cern.ch/record/1460438?ln=en;\\
ATLAS-CONF-2012-093, \\
http://cdsweb.cern.ch/record/1460439.

\bibitem{CMSSMscan} 
O.~Buchmueller {\it et al.},
arXiv:1112.3564;
%
D.~Ghosh, M.~Guchait, S.~Raychaudhuri and D.~Sengupta,
arXiv:1205.2283;
%
C.~Balazs, A.~Buckley, D.~Carter, B.~Farmer and M.~White,
arXiv:1205.1568;
%
P.~Bechtle {\it et al.},
arXiv:1204.4199;
%
H.~Baer, V.~Barger and A.~Mustafayev,
arXiv:1202.4038;
%
S.~Akula, P.~Nath and G.~Peim,
arXiv:1207.1839;
%
H.~Baer, V.~Barger, P.~Huang, A.~Mustafayev and X.~Tata,
arXiv:1207.3343;
%
J.~Cao, Z.~Heng, J.~M.~Yang and J.~Zhu, 
arXiv:1207.3698.



\end{thebibliography}
\end{document}